\documentclass[letterpaper,twocolumn,10pt]{article}
\usepackage{usenix2019,epsfig,endnotes}
\usepackage[sort, numbers]{natbib}
\usepackage{tikz}
\usepackage{amsmath}
\usepackage{amsfonts}
\usepackage{xcolor, colortbl}
\usepackage{collcell}
\usetikzlibrary{positioning}
\usepackage[figure]{algorithm2e}
\usepackage{pgfplots}
\usepackage{booktabs}
\usepackage{subfig}
\usepackage{tabularx}
\usepackage{diagbox}
\usepackage{multirow}
\usepackage{caption}
\usepackage{enumitem}
\usepackage{bigdelim}
\usepackage{calc}
\usepackage{siunitx}
\usepackage{fancyhdr}
\usepackage{hyperref}
\usepackage[noabbrev]{cleveref}
\usepackage{diagbox}
\usepackage{graphicx}
\usepackage{bm}

\crefrangelabelformat{Figure}{#3#1#4--#5\crefstripprefix{#1}{#2}#6}

\newcommand{\code}[1]{{\small\texttt{#1}}}

\newlength\figureheight
\newlength\figurewidth
\setlist{topsep=0pt,nolistsep}

\usepackage{setspace}

\begin{document}

%don't want date printed
\date{}

%make title bold and 14 pt font (Latex default is non-bold, 16 pt)
\title{\Large \bf Adversarial Binaries for Authorship Identification}

%for single author (just remove % characters)
\author{
Xiaozhu Meng, Barton P. Miller, and Somesh Jha \\
Computer Sciences Department \\
University of Wisconsin - Madison \\
Madison, WI, USA, 53706 \\
\{xmeng,bart,jha\}@cs.wisc.edu \\
} % end author

\maketitle

% Use the following at camera-ready time to suppress page numbers.
% Comment it out when you first submit the paper for review.
%\thispagestyle{empty}

\subsection*{Abstract}
Binary code authorship identification determines authors of a binary program. Existing techniques have used supervised machine learning for this task. In this paper, we look at this problem from an attacker's perspective. We aim to modify a test binary, such that it not only causes misprediction but also maintains the functionality of the original input binary. Attacks against binary code are intrinsically more difficult than attacks against domains such as computer vision, where attackers can change each pixel of the input image independently and still maintain a valid image. For binary code, even flipping one bit of a binary may cause the binary to be invalid, to crash at the run-time, or to lose the original functionality. 

We investigate two types of attacks: untargeted attacks, causing misprediction to any of the incorrect authors, and targeted attacks, causing misprediction to a specific one among the incorrect authors. We develop two key attack capabilities: feature vector modification, generating an adversarial feature vector that both corresponds to a real binary and causes the required misprediction, and input binary modification, modifying the input binary to match the adversarial feature vector while maintaining the functionality of the input binary.

We evaluated our attack against classifiers trained with a state-of-the-art method for authorship attribution. The classifiers for authorship identification have 91\% accuracy on average. Our untargeted attack has a 96\% success rate on average, showing that we can effectively suppress authorship signal. Our targeted attack has a 46\% success rate on average, showing that it is possible, but significantly more difficult to impersonate a specific programmer's style. Our attack reveals that existing binary code authorship identification techniques rely on code features that are easy to modify, and thus are vulnerable to attacks.

\section{Introduction}

The task of binary code authorship attribution is 
to determine the authors of a binary program,
and has significant application to malware forensics,
software supply chain risk management, and software plagiarism detection.
Recent studies \cite{caliskan2018coding, Alrabaee2014Authorship, Rosenblum2011BinAuthor, Meng2017MultiAuthor, Meng2018MultiToolchain}
have made significant progress in developing
machine learning based techniques to identify authors of binary programs.
In this paper, we look at the problem of authorship identification
from an attacker's perspective and attempt to perform authorship evasion,
whose goal is to trick machine learning classifiers for authorship identification
into making wrong predictions.
We show that adversarial machine learning can pose a threat
to binary code authorship identification 
when confronted with a carefully crafted binary code artifact,
causing these classifiers to produce misleading results.

Authorship evasion is the application of adversarial machine learning
to authorship identification.
The field of adversarial machine learning
has focused on attacking and defending machine learning 
systems used in real world applications \cite{biggio2017wild}.
A specific threat is called a test time attack, 
where attackers change a test example 
to cause misprediction.
Researchers have performed successful test time attacks for a wide range of domains,
including computer vision \cite{Biggio2013EAA, carlini2016towards,szegedy2013intriguing}, 
audio processing \cite{zhang2017dolphinattack}, and program analysis tasks \cite{Grosse2017AEM, simko2018recognizing}. 
Such test time attacks can have serious security implications.
For example, 
Grosse et al. \cite{Grosse2017AEM} showed that they can change the manifest file of an Android program
to circumvent malware detection;
and Simko et al. \cite{simko2018recognizing} showed that
when given source code from other people, 
a programmer can change the source code
to avoid authorship attribution.

However, currently there are no such attacks against binary code authorship identification.
The key challenge for developing such attacks is to 
modify the binary to not only cause misprediction,
but also maintain the structural validity and functionality of the binary.
Even flipping one bit of a binary may cause the binary 
to either be invalid, such as not loadable by the loader,
or lose functionality that the attackers care about. 
Therefore, 
attacks against binary code are intrinsically more difficult than attacks
targeted at domains such as computer vision,
where attackers can change each pixel of the input image
independently and still maintain a valid image.

\begin{figure*}
	\small
	\centering
	\includegraphics[page=1, trim={0in, 1.5in 3.3in 0in}, clip, height=3in]{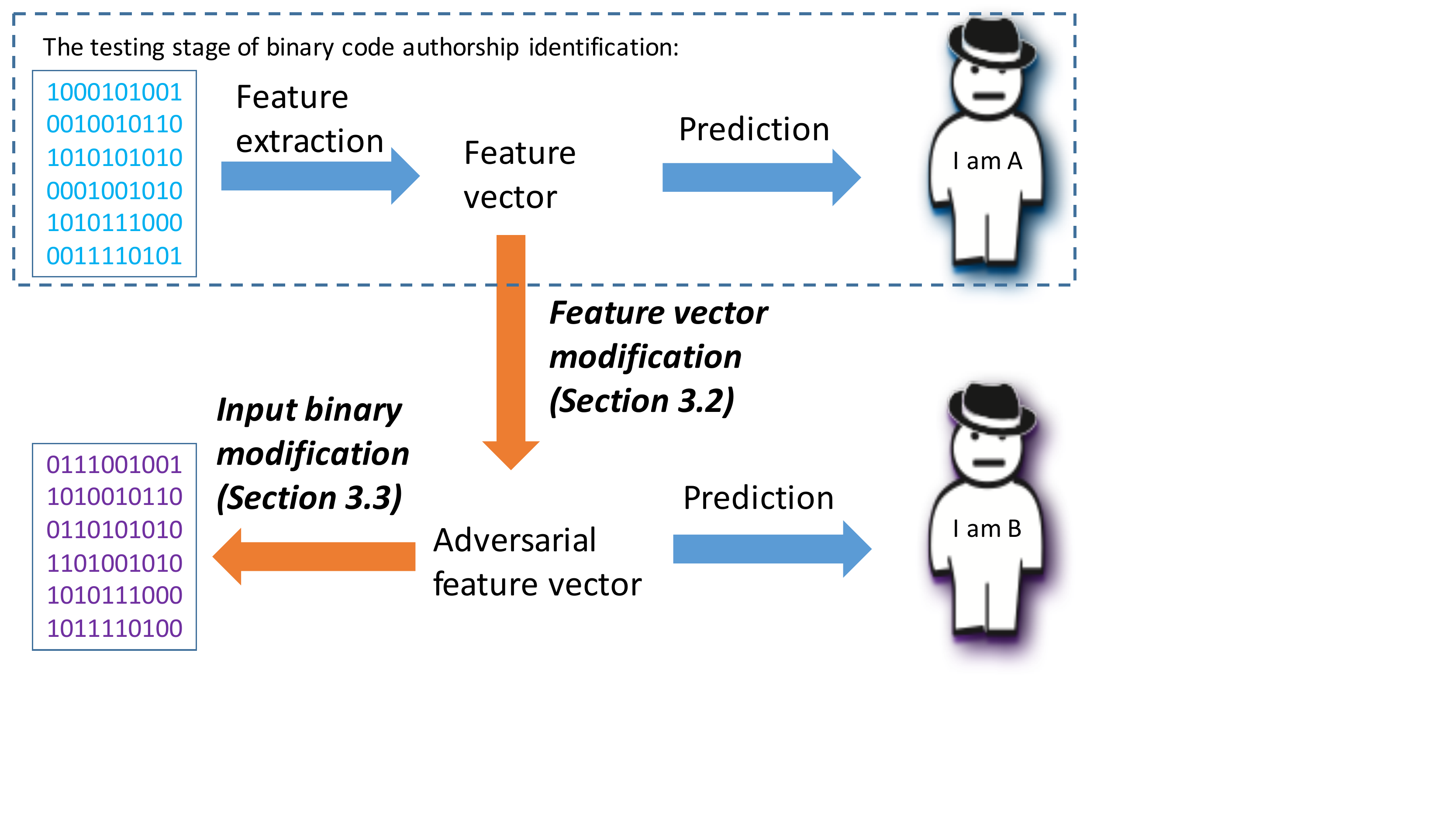}
	\caption{\textbf{Overview of our attack.} Our attack includes two key steps.
	Feature vector modification generates an adversarial feature vector
	that corresponds to a real binary and causes the required misprediction.
	Input binary modification generates a new binary that matches the generated
	adversarial feature vector.
	The dashed blue box illustrates the testing stage of binary code authorship identification.
	A blue arrow represents a step from the testing stage.
	An orange arrow represents a step of our attack.
	}
	\label{fig:overview}
\end{figure*}

In this paper, we present a framework for automatically attacking 
techniques for binary code authorship identification.
The contributions of our attack framework are three-fold.

\begin{enumerate}
\item We show that it is realistic to 
automatically attack binary in an end-to-end fashion:
we take a binary program as input and generate a new, valid binary
that has the same functionality as the input binary and
causes misprediction. We believe that our attack framework can be used
for attacking other binary code forensic tasks, such as malware detection
and compiler identification.

\item Our techniques can be used for adversarial re-training to train more secure classifiers,
incorporating the generated adversarial examples into the training set
to re-train a classifier with a modified loss function \cite{miyato2015distributional,madry2018towards}.

\item We reveal the weaknesses of a state-of-the-art technique for
binary code authorship identification \cite{caliskan2018coding}
and summarize the lessons we learned for designing more secure machine learning systems for 
binary analysis tasks. 
We stress that it is important to assess learning based techniques from
the attacker's perspective.
\end{enumerate}

We focus on two different types of attacks:
the \textit{untargeted attack}, which is to cause misprediction
to any of the incorrect authors;
and the \textit{targeted attack}, which is to cause misprediction
to a specific incorrect author.

Stealthiness is an important design goal of our attack.
Our attack should not leave obvious footprints that can be easily detected.
We aim to improve stealthiness in two dimensions.
First, the generated adversarial binary should 
be similar to the original binary in structure.
We prefer small and local modifications
over large and global modifications.
Second, our attack should be diversified, meaning
when running multiple times with the same input, 
our attack should generate different adversarial binaries.
Diversified attacks make hash-based detection strategies ineffective. 

We make two main assumptions about the threat model.
First, the attackers have perfect knowledge of target authorship identification tool.
This assumption allows performing a worst-case evaluation of the security
of the target authorship identification tool,
common when performing test time attacks 
\cite{Biggio2013EAA, carlini2016towards,szegedy2013intriguing,simko2018recognizing}.
Second, the attackers plan to perform a test time attack,
so they can affect the prediction results
only by providing a crafted input binary.
Other possible attacks against learning systems 
such as training set poisoning \cite{Biggio2012PAA,mei2015using}
are not in the scope of this paper.

Authorship identification techniques have a training stage and a testing stage.
While we do not directly attack the training stage,
three choices made in this stage impact our attacks.
First, the design of the binary code features
determines the program properties of the binary 
to modify during attacks.
Features are typically defined to describe program properties 
including machine instructions, program control flow, 
constant strings, and program meta-data such as function symbols.
Second, identification techniques use binary code analysis tools such as 
Dyninst \cite{DyninstAPI}, NDISASM \cite{NDISASM} or Radare2 \cite{radare2}
for feature extraction.
A key part of our attack is to modify the binary and trick 
the binary code analysis tools into extracting modified features to cause misprediction.
Third, based on the machine learning algorithm used by the identification technique,
the attacker may need to use different attack algorithms
to determine which features should be modified to cause misprediction.
There are existing attack algorithms for a variety of learning models,
including
Deep Neural Networks (DNNs) \cite{carlini2016towards},
Random Forests (RFs) \cite{Kantchelian2016EHT},
and Support Vector Machines (SVMs)\cite{Grosse2017AEM,Biggio2013EAA}.

Our attack ties closely to the testing stage.
Figure~\ref{fig:overview} illustrates the testing stage
and the key steps of our attack.
The testing stage has two key steps:
extracting code features from the input binary to
construct a feature vector and
applying the pre-trained model on the feature vector to generate the prediction results.
Our new attack focuses on developing two interacting attacking abilities:
\textit{feature vector modification}, 
generating an adversarial feature vector that corresponds to a real binary
and causes the required misprediction,
and \textit{input binary modification}, 
modifying the input binary to match the adversarial
feature vector while \textit{maintaining the functionality of the input binary}. 

Our attack framework introduces a large space for generating diversified attacks.
Given an input binary and the misprediction target,
feature vector modification can generate different adversarial feature vectors 
to cause the required misprediction.
Given an adversarial feature vector, 
input binary modification can generate different adversarial binaries
to match the feature vector.

Our basic idea of feature vector modification is to integrate
multiple modification strategies to increase the chance
of finding the desired feature vector and to generate diversified feature vectors.
Candidate strategies include random feature modification,
the attack presented by Carlini and Wagner \cite{carlini2016towards} (denoted as the CW attack),
and the \textit{projected gradient descent} (PGD) \cite{ilyas2018prior}.
We extend these modification strategies in two ways to 
address the structural validity requirement of binary programs:

First, existing attacks modified each feature independently; 
changing one pixel of an image does not impact other pixels. 
However, features in our domain can be correlated. 
Without considering feature correlation,
we may generate feature vectors
for which there do not exist corresponding valid binaries.

We perform a feature correlation analysis 
to derive feature correlation from a substitute data set.
Note that this data set can be, but does not have to be 
the training set used for training the target classifier.
We can derive useful feature correlation information,
as long as this data set is drawn from the same application domain as the training set.
We then use the correlation information to ensure
that correlated features are modified in a consistent way.

Second, existing attacks did not consider the difficulty of modifying a feature;
changing any pixel of an image is equally easy for maintaining the validity of the image.
However, for binary analysis, 
some features are easier to modify than others.
For example, local features that describe machine instructions
are typically easier to modify than global features
that describe program control flow, because modifying global features
can require changing more code, making it more difficult to maintain structural validity.

We categorize binary code features into a small number of feature groups
such that features in a group can be modified with the same strategy.
We attempt to modify one feature group at a time until causing misprediction.
Grouping features also allows generating diversified adversarial feature vectors
by modifying different combinations of feature groups.

Our input binary modification removes or injects features 
according to the results of feature vector modification,
with the additional goals of maintaining structural validity and preserving functionality.
To remove features, we need to ensure that
the program properties that correspond to the removed feature 
are replaced with semantically equivalent ones.
In many cases, we cannot simply remove them because such modification
would break the functionality of the binary.
On the other hand, 
the main challenge of injecting features 
is to ensure that the binary code analysis tools used for feature extraction
indeed recognize the injected code, data or meta-data.

We observe that the space of binary modification is large
and there could be many different binaries matching the given adversarial feature vector.
Therefore, we show the feasibility of our attack by construction.
We design injection and removal strategies for each feature group.
These modification strategies consist of 
a sequence of binary modification primitives,
including inserting, deleting, and replacing
code, data, and meta-data.
Our modification primitives use randomization,
thus add another dimension to
the diversity of our attack.
Binary instrumentation and rewriting tools such as Dyninst support
the implementation of these modification primitives.

We evaluate our evasion attacks
using five classifiers
trained with the techniques presented by Caliskan-Islam et al. \cite{caliskan2018coding}.
We achieved 96\% success rate for untargeted attacks
and 46\% success rate for targeted attacks.
Our results show that we can effectively suppress authorship signal
for authorship evasion, but it is significantly more difficult
to impersonate the style of another author.
Our results also reveal the weakness in current authorship identification techniques.
Many features used in current authorship identification techniques 
are based on program properties that are easy to manipulate.
We can automatically modify these features, 
making such classifiers vulnerable to test time attacks.

\section{An Attack Example}
\label{sec:evasion:example}
We present an example showing 
how to perform untargeted attacks to a classifier 
for binary code authorship attribution.
The goal of this section is to give an overview 
of our attack process.
In the subsequence sections,
we describe the steps in more details.

We first describe the procedures for setting up the target classifier,
which is trained with the techniques presented by Caliskan-Islam et al. \cite{caliskan2018coding}.
We then describe how to generate feature vectors that correspond to real binaries and cause misprediction.
Finally, we give examples on how to modify the binary to match the generated feature vectors.

\subsection{Binary Code Authorship Attribution}
\label{sec:example:background}

Caliskan-Islam et al. \cite{caliskan2018coding}
assume that a binary is written by a single author,
so, they predict one author for a binary.
Their workflow can be summarized in four steps.

\begin{enumerate}
\item \textit{Define candidate features}:
They used binary code features that describe machine instructions 
and program control flow.
They also included source code features derived from decompiled source code.
The source code features include character n-grams and tree n-grams. 
The tree n-grams are extracted from abstract syntax trees (ASTs) built by parsing the source code.
These source code features have been shown to be effective 
for source code authorship attribution \cite{Caliskan2015SourceCodeAuthorship}.

\item \textit{Extract features}:
They used two disassemblers, 
NDISASM \cite{NDISASM} and radare2 \cite{radare2},
to extract binary code features. 
To derive source code features,
they first used the Hex-Ray decompiler \cite{HexRays},
and then used Joern \cite{Yamaguchi2014MDV} to parse the source code into ASTs.
They represent each feature as a string.
To derive feature strings, 
they first split the results of disassembly, decompiling, and source code parsing
into tokens and then normalize hex tokens to the generic symbol ``hexdecimal''
and decimal digit tokens to the generic symbol ``number''.
They use string matching to count the frequency of a feature string and
use the frequencies of feature strings to construct feature vectors.

\item \textit{Select Features}: 
Typically, hundreds of thousands of features are extracted from a data set.
So, feature selection is necessary to avoid overfitting.
They selected features that have information gain with respect to the author labels.

\item \textit{Train a classifier}: 
They compared Random Forests (RFs) with Support Vector Machines (SVMs)
and reported that RFs outperformed SVMs.
\end{enumerate}

They used a data set derived from Google Code Jam (GCJ)
and evaluated their techniques with binaries compiled by GCC on a 32-bit platform. 
For binaries compiled with GCC and \code{-O0},
they achieved 96\% accuracy for classifying 100 authors.
For binaries compiled with higher optimization levels, 
they reported slightly lower accuracy.

We obtained the GCJ source files used by Caliskan-Islam et al. \cite{caliskan2018coding} 
and their source code for extracting features.
Due to the predominance of 64-bit platforms,
we perform attacks on 64-bit platforms.
Note that while Caliskan-Islam et al. only evaluated their techniques on 32-bit platforms,
their techniques can be directly applied to 64-bit platforms.
We compiled the GCJ sources with GCC 5.4.0, using \code{-O0} optimization on a 64-bit platform,
and achieved 90\% accuracy for classifying 30 authors.

\subsection{Feature Vector Modification}
\label{sec:example:proc}

Given the target classifier to attack, 
the goal of our attack is to modify an input binary to cause the required misprediction.
The two key steps for attacking this authorship attribution classifier
are generating feature vectors that can correspond to a real binary and cause the required misprediction,
and modifying the input binary to match the feature vector.
We use examples to illustrate the importance of 
our feature correlation analysis and feature grouping 
on generating an adversarial feature vector.

\subsubsection{Feature Correlation Analysis}

We derive correlations between features
to guide feature vector modification to 
generate feature vectors corresponding to real binaries.

We identify two types of feature correlation for this classifier.
First, a feature can contain other features.
For example, if feature ``\code{push rax; push rbx}'' is present in a binary,
features ``\code{push rax}'' and ``\code{push rbx}'' are also present.
So, the frequencies of ``\code{push rax}'' and ``\code{push rbx}'' 
should be no fewer than the frequency of ``\code{push rax; push rbx}''.
Second, the same properties extracted by different binary analysis tools 
are treated as different features.
For example, the instruction ``\code{call fprintf}'' corresponds
to three different features:
``\code{call fprintf}'' extracted by NDISASM, 
``\code{call fprintf}'' extracted by radare2, 
and ``\code{fprintf}'' extracted from decompiled source code.
These features should all have the same frequency. 

We derive linear correlation between features based on the training set.
For each pair of features, we perform linear regression and calculate the correlation coefficient.
If the coefficient is larger than a threshold value, such as $0.9$,
we merge the pair into one feature.
While this simple strategy will miss non-linear feature correlation,
our experiments showed that capturing linear correlation is sufficient 
for launching successful attacks against authorship attribution.

\subsubsection{Generating Adversarial Feature Vectors}

Our framework supports multiple feature vector modification strategies,
including random modification, 
the attack presented by Carlini and Wagner \cite{carlini2016towards} (denoted as the CW attack),
and the \textit{projected gradient descent} (PGD).
We use the CW attack in this example as it out-performs other attacks in our experiments.
The CW attack is designed for DNNs trained for images,
and can be readily applied to other gradient based learning algorithms.
However, Caliskan-Islam et al. used RFs, which is a non-gradient based learning algorithm.
Fortunately, 
researchers have shown that adversarial examples created for classifiers trained with 
one type of learning algorithms (such as DNN)
are likely to cause misprediction for classifier trained with 
a different type of learning algorithms (such as RF) \cite{papernot2016transferability,TPGB17}.
Therefore, we first trained a substitute DNN using the same training data
and then applied the adversarial vectors to the RF classifier.
The substitute DNN is a simple feed-forward neural network,
containing 7 hidden layers with each layer having 50 hidden units.
The substitute DNN has 80\% accuracy.
While the substitute DNN has modestly lower accuracy than the target classifier,
as we will show in Section~\ref{sec:evasion:eval},
this accuracy gap does not impact the success rate of our attack.

To ensure that the generated adversarial feature vector confuses not only the substitute classifier
but also the target classifier, we keep generating new feature vectors
until the resulting vectors can mislead the target classifier.
Our new attack strategy can generate effective adversarial feature vectors,
reducing the accuracy of both the substitute DNN and the RF classifier to 0\%.

However, it is difficult to modify the input binary 
to completely match the feature vectors generated in this way,
as they contain hundreds of modified features.

\subsubsection{Categorizing Features}

We have observed that 
while the attacks presented by Carlini and Wagner 
can make effective changes to the feature vector to cause misprediction,
not all changes are necessary for causing misprediction.
Therefore, we attempt to modify fewer features to cause misprediction,
making it easier to perform binary modification to match the generated feature vector.
We categorize features into feature groups,
so that features in the same feature group can be modified with the same strategy.
And then we modify one feature group at a time until misprediction occurs.

Two important factors for categorizing the features 
are the program properties that the features describe
and the strength of the binary analysis tools.
For the first factor,
features describing low level code properties such as machine instructions
are easier to modify compared to features describing higher level structural properties such as
program control flow and data flow.
Therefore, we started by attacking instruction features.

For the second factor,
recall that Caliskan-Islam et al. used two disassemblers:
NDISASM, which disassembles the binary linearly from the first byte of the binary file,
and radare2, which understands the layout of the binary, performs binary analysis
to identify code bytes, and attempts to disassemble only code bytes.
It is easier to modify features extracted by NDISASM, 
because NDISASM also disassembles non-loadable sections
and editing or adding non-loadable sections has no impact on 
the functionality of the program.
On the other hand, 
instruction features extracted by radare2 typically represent real code.
So, we need to ensure that we do not change the functionality when removing a radare2 feature,
and ensure that radare2 disassembles the inserted code when injecting a radare2 feature.

After grouping features, we first modify instruction features extracted by NDISASM,
reducing the accuracy from 90\% to 45\%.
We then modify instruction features extracted by radare2,
further reducing the accuracy from 45\% to 7\%.
Note that only features in the these two feature groups are modified
and we can generate new binaries to complete the attack.

\subsection{Binary Modification Strategies}

Finally, we describe our binary modification strategies for 
injecting and removing NDISASM and radare2 features, using four typical examples.
These examples are extracted from our successful attacks.
In each example, we describe the modification primitives that constitute the modification strategy and 
explain why our modifications do not change the functionality of the input binary. 

\subsubsection{Modifying NDISASM Features}

\begin{figure}
   \small
   \centering
   \begin{tabular}{l p{5cm}}
Feature string & \code{or [rax],ebp}  \\
Raw bytes & \code{09 28} \\
\hline
Modification & \small{Insert bytes \code{09 28} into a new non-loadable section} \\
  \end{tabular}
   \caption[An example of injecting a NDISASM feature]
   {\small \textbf{An example of injecting a NDISASM feature}. We can insert the bytes
   into a non-loadable section.}
   \label{fig:NDISASM:feat1}
\end{figure}

We show two examples of modifying NDISASM features. 
The first example shows the case where we can inject a feature by inserting
bytes into the binary.                             
As shown in Figure~\ref{fig:NDISASM:feat1},
we need to inject instruction feature ``\code{or [rax],ebp}'' into the target binary.
Since NDISASM disassembles every bytes in the binary,
we can add a new non-loadable section to store the bytes of the corresponding instruction.
This simple injecting strategy 
causes NDISASM to extract this feature 
and does not change the functionality of the program.

\begin{figure}
\small
   \centering
   \begin{tabular}{l p{5cm}}
Binary name & 1835486\_1481492\_paladin8 \\
Feature string & \code{imul ebp,[fs:rsi+hexadecimal], dword hexadecimal} \\
Offset in the binary & 0x3e09 \\
Raw bytes & \code{64 69 6E 38 2E 63 70 70} \\
\hline
Modification & \small{Overwrite bytes to other values} \\
   \end{tabular}
   \caption[An example of removing a NDISASM feature]{\small \textbf{An example of removing a NDISASM feature.} 
   This feature seems to represent an instruction, but actually represents a string in the \code{.strtab} section.}
   \label{fig:NDISASM:feat2}
\end{figure}

The second example shows the case where we can simply remove a feature, 
without replacing the removed program property with a semantically equivalent one.
As shown in Figure~\ref{fig:NDISASM:feat2},
this feature seems to represent an \code{imul} instruction.
However, offset 0x3e09 of the binary is in the \code{.strtab} section,
which stores symbol names for the compile-time symbol table.
Therefore, instead of representing an instruction, 
the feature represents string ``in8.cpp''.
To remove this feature,
We can change the string ``in8.cpp'' to any another string.
\code{.strtab} is used at debug-time, and not used at the link-time or run-time
(it disappears if the binary is stripped), so
changing its content does not impact the functionality of the original program.

In addition, we tried to understand why the string ``in8.cpp'' is a useful feature.
We found that the string is extracted from source file name ``1835486\_1481492\_paladin8.cpp''
and ``paladin8'' is the author's name.
So, this feature turns out to contain three characters of the author's name.
While a string containing three characters of the author's name
is useful for identifying the author,
such author name feature is not available in any realistic context.
This example teaches us a lesson that
machine learning practitioners need to ensure that 
the feature definition and the extracted features actually match.
In this case, instruction features should only be extracted from real code bytes.
So, the use of NDISASM is not robust for real world identification
because it disassembles all bytes in the binary.

\subsubsection{Modifying radare2 Features}

We now show two examples of modifying radare2 features.
The first example shows the case where we need to insert new code and data.
As shown in Figure~\ref{fig:radare2:feat1},
feature ``number.in'' represents a string.
Note that this feature is not present in the target binary,
and we need to inject it into the target binary to cause misprediction.
We found feature ``number.in'' in another binary,
based on the instruction ``\code{mov \$0x400c57,\%edi}''.
Here, address 0x400c57 points to a string ``number.in'';
radare2 recognizes the string and prints it in the disassembly results.

To inject this feature, 
we need to (1) insert string ``number.in'' into the target binary,
and (2) insert a \code{mov} instruction that loads the address of the inserted string. 
However, to trick radare2 to disassemble the inserted instruction,
there are two additional steps.
First, we create a function symbol pointing to the inserted code.
Second, we append a return instruction after the inserted code.
Since most binary analysis tools treat function symbols as ground truth
for specifying the locations of code bytes,
our injection strategies can be also applied to other binary analysis tools.

\begin{figure}
\small
   \centering
   \begin{tabular}{l p{5cm}}
Feature string & \code{number.in} \\
Machine instruction & \code{mov \$0x400c57,\%edi} \\
\hline
                    &  \small{1. Insert string ``number.in'' into a new data section} \\
\multirow{-2}{*}{Modifications} & \small{2. Insert new instructions to load the inserted string} \\                     
  \end{tabular}
   \caption[An example of injecting a radare2 feature]
   {\small \textbf{An example of injecting a radare2 feature}. This feature represents a string.
   We need to insert the string and insert an instruction to load the address of the string.}
   \label{fig:radare2:feat1}
\end{figure}

The second example shows the case where we need to replace existing code
with semantically equivalent code to remove a feature.
As shown in Figure~\ref{fig:radare2:feat2}, we need to remove a feature describing an object symbol.
The feature is extracted from instruction ``\code{mov 0x20157d(\%rip),\%rax}''.
Here radare2 recognizes that the result of 
the PC-relative calculation points to an object symbol,
so it annotates the instruction with the name of the object symbol in the disassembly results.

To remove this feature, we need to transform the calculation of the symbol address 
to a semantically equivalent calculation done by one or more instructions,
so that radare2 cannot recognize the loading of the symbol address.
To do this, we can split the address loading into two instructions: 
loading the address minus one into the target register
and incrementing the target register by one.
We cannot just overwrite the symbol name with a different string because
this symbol is in the \code{.dynsym} section and it is used for dynamic linking
(Overwriting the name of a dynamic symbol will cause the program to not be loadable).

\begin{figure}
\small
   \centering
   \begin{tabular}{l l}
Feature string & \code{obj.stdin} \\
Machine instruction & \code{mov 0x20157d(\%rip),\%rax} \\
\hline
          & \small{1. Load }\code{0x20157d(\%rip)-1}\small{ into }\code{\%rax} \\
\multirow{-2}{*}{Modifications} & \small{2. Increment }\code{\%rax} \\         
  \end{tabular}
   \caption[An example of removing a radare2 feature]
   {\small \textbf{An example of removing a radare2 feature.} 
   This feature represents an object symbol.
   We can split the address loading instruction into two instructions to remove the feature.}
   \label{fig:radare2:feat2}
\end{figure}
\section{Attack Framework}
\label{sec:evasion:framework}

\LinesNumberedHidden
\begin{algorithm}[t]
        %\scriptsize
        \SetAlgoNoEnd
        \SetInd{0.5em}{0.5em}
        \SetKwInOut{Input}{input}
        \SetKwInOut{Output}{output}
                \Input{an input binary $b$; a pre-trained model $m$; feature groups $fg$; and a misprediction target $tar$ ($tar=-1$ represents untargeted attacks)}
                \Output{an adversarial binary $b'$ that causes misprediction}
                \vspace{0.5em}
                \ShowLn $P$ $\leftarrow$ \code{FeatureCorrelationAnalysis}($m$)\;
                \ShowLn $\bm{x}$ $\leftarrow$ \code{FeatureExtraction}($b$)\;
                \ShowLn $y$ $\leftarrow$ \code{Prediction}($m$, $\bm{x}$)\;
                \tcp{Keep looping until causing misprediction}              
                \ShowLn\For{$g$ in $fg$} {
                        \ShowLn $\bm{x'}$ $\leftarrow$ \code{FeatureVectorModification}($\bm{x}$, $g$, $P$)\;
                        \ShowLn $b'$ $\leftarrow$ \code{InputBinaryModification}($b$, $\bm{x'}$, $g$, $P$)\;
                        \ShowLn $y$ $\leftarrow$ \code{Prediction}($m$, \code{FeatureExtraction}($b'$))\;
                        \tcp{Non-targeted attacks succeed}
                        \ShowLn \lIf{$tar == -1$  \textbf{and} $y \neq y'$}{\textbf{break}}
                        \tcp{Targeted attacks succeed}
                        \ShowLn \lIf{$tar \neq -1$  \textbf{and} $tar == y'$}{\textbf{break}}
                }                
        \caption[Our attack algorithm]{\small\textbf{The attack algorithm.}
        The main structure of the algorithm is to iterate over feature groups
        until we generate a new binary that causes misprediction.}
        \label{fig:framework}
\end{algorithm}

We describe our attack framework in this section,
based on the attack algorithm in Figure~\ref{fig:framework}.
The inputs to our algorithm includes an input binary $b$, 
a target classifier $m$, 
feature groups $fg$,
and a misprediction target label $tar$.
The output of the algorithm is an adversarial binary $b'$ that causes the required misprediction.
The main component of our algorithm is an attack-verify loop,
where we iterate over feature groups until we generate
a new binary that causes misprediction.

Our algorithm relies on two routines from the machine learning application we are attacking:
\code{FeatureExtraction} to extract features
and \code{Prediction} to generate a prediction label from a set of known labels.
The meaning of these labels depend on the target application.
For example, a label can describe an author for authorship attribution
or a compiler for compiler identification.
We now describe the other routines in our algorithm.

\subsection{Feature Correlation Analysis}
\label{sec:framework:corr}

Given a set of features $F=\{f_1, f_2, \ldots, f_k\}$ used in the target classifier $m$,
our feature correlation analysis
generates a partitioning of the features, $P=\{p_1, p_2, \ldots, p_k\}$, 
where each partition consists of all correlated features.
So, $\forall f_x\in p_i$ and $f_y\in p_i$, $f_x$ and $f_y$ are correlated;
and $\forall i \neq j$, $f_x\in p_i$, and $f_y\in p_j$, 
$f_x$ and $f_y$ are not correlated.
In addition, feature partitions are disjoint.
So, $\forall i \neq j$, $p_i \cap p_j = \emptyset$. 

We build a undirected graph to generate the feature partitioning.
Let $G=(V,E)$, where each node in the graph represents a feature (so $V=F$),
and each edge in the graph represents the correlation between two features.
We only capture linear correlation between features,
creating an edge between two nodes 
if the linear correlation coefficient between two features
is larger than a pre-specified threshold. 
In another words, $E=\{(f_i, f_j): coe_{ij} \geq T \}$, 
where $coe_{ij}$ is the linear correlation coefficient between $f_i$ and $f_j$
and $T$ is the pre-specified threshold.
Finally, each connected component in the graph represents a partition of the correlated features.

An important observation is that
we do not have to capture the exact correlation between features
to launch successful attacks. 
For example, suppose we have three features:
``$f_1$: push rax; push rbx'', ``$f_2$: push rax'', and ``$f_3$: push rbx''.
The precise correlation is

\begin{equation} \label{cor:1} 
(freq(f_1)\leq freq(f_2)) \wedge (freq(f_1)\leq freq(f_3))
\end{equation}
where $freq(f)$ represents the frequency of feature $f$.
Our algorithm will put all three features in the same partition
and derive the following correlation:
\begin{equation} \label{cor:2}
freq(f_2)=A_1freq(f_1)+B_1, freq(f_3)=A_2freq(f_1)+B_2
\end{equation} 

As we will discuss in the next section,
it is straightforward to incorporate correlation (\ref{cor:2}) into our feature vector modification.
In addition, 
as the linear correlation is derived from a data set drawn from the same domain as the training set for the target classifier,
feature vectors satisfying correlation (\ref{cor:2}) typically also satisfy correlation (\ref{cor:1}).

\subsection{Feature Vector Modification}
\label{sec:framework:vec}

Given an input feature vector $\bm{x}=[x_1, x_2, \ldots, x_k]$, 
where $x_i$ represents the feature value of feature $f_i$,
our feature vector modification outputs a modified feature vector $\bm{x'}$,
such that the prediction results for $\bm{x'}$ are 
different from the prediction results of $\bm{x}$ (for untargeted attacks)
or are the specified results (for targeted attacks).

Our attack framework supports multiple strategies for generating 
the required feature vector, including random modification, 
the attack presented by Carlini and Wagner \cite{carlini2016towards} (denoted as the CW attack),
and the \textit{projected gradient descent} (PGD).
The basic idea of these strategies is to define $\bm{x'}$ as $\bm{x}+\bm{\delta}$; so
once we have calculated $\bm{\delta}$, we know $\bm{x'}$.
They differ in terms of how to calculate $\bm{\delta}$.
Adversarial learning based strategies such as the CW attack 
convert the calculation of $\bm{\delta}$ into numerical optimization problems
and utilize general purpose optimizer such as Adam \cite{kingma2014adam}.

We make three modifications to the CW attack and the PGD attack when applying
them to our domain.
First, they may generate feature vectors with non-integer values. 
However, as discussed in Section~\ref{sec:example:background}, 
Caliskan-Islam et al. \cite{caliskan2018coding}
used feature counts to construct feature vectors.
So, $\bm{x'}$ should only have integer values.
A simple strategy that works well for us is to round values generated by
the CW attack to the nearest integer.

Second, we must incorporate the feature correlation information
derived in Section~\ref{sec:framework:corr} into the attack.
To do this,
we normalize each individual feature to a Gaussian with zero mean and unit variance,
merge all correlated features into one feature, and let the CW attacks work with only the merged features.
Recall that we track linear correlation between features;
for two correlated features $f_1$ and $f_2$, $freq(f_1)=Afreq(f_2)+B$.
After the normalization step, $A$ is normalized to $1$ and $B$ is normalized to $0$.
Therefore, we can merge them into a single feature.

Third, we found that existing attacks often did not generate 
an adversarial feature vectors with the minimal number of modified features.
So, we design a two-step post-processing to further 
reduce the number of modified features and the magnitude of changes.
First, for each modified feature, we undo the modification and set its value to its unmodified value. 
If we can still cause misprediction, we finalize the undo of the modification.
Second, for each modified feature, we enumerate every integer between the unmodified value
and the new value. We set the value of this feature 
to the one that is closest to the unmodified value 
and causes misprediction. 

\subsection{Binary modification strategies}
\label{sec:framework:modi}

Given a new feature vector $\bm{x'}$ that causes misprediction,
we describe how to modify the input binary to match $\bm{x'}$,
grouping features based on the program properties that the features describe
and the binary analysis tool used to extract the feature.
We also describe feature injection and removal strategies for feature groups.
Our modification strategies
consist of binary modification primitives supported by tools such as Dyninst \cite{DyninstAPI}.
Finally, we discuss how to determine which modification strategy to use for a specified feature
and how to generate diverse adversarial binaries.

\subsubsection{Feature injection strategies}

\begin{table*}[t!]
\small
    \centering
    \caption[Summary of feature injection strategies]{\small \textbf{Summary of feature injection strategies}.
    The first column lists the program properties to inject.
    The second and third columns list the binary modification primitives
    used for the injection.}
    \begin{tabular}{ l | c | c } %p{100mm} |}
    \cellcolor{gray!20}\textbf{Program Property}   &   \cellcolor{gray!20}\textbf{NDISASM}     &    \cellcolor{gray!20}\textbf{radare2} \\
    \hline
    Instructions \code{I} &  \code{InsertNonCodeBytes(I)} &  \code{InsertFunction(I)} \\
    \hline
    \multirow{2}{*}{Loading symbol \code{S}}    &  \multirow{2}{*}{NA}    &  \code{addr = InsertSymbol(S)} \\
                                                  &                         &  \code{InsertFunction(loading addr)} \\
    \hline
    \multirow{2}{*}{Loading data  \code{D}}   &  \multirow{2}{*}{NA}     &  \code{addr = InsertData(D)} \\
                                                   &                          &  \code{InsertFunction(loading addr)} \\  
    \hline
    \multirow{2}{*}{Calling function \code{F}}   &  \multirow{2}{*}{NA}     &  \code{addr = InsertCall(F)} \\
                                                   &                          &  \code{InsertFunction(calling addr)} \\        

    \end{tabular}
    \label{table:featinjection}
\end{table*}

Table~\ref{table:featinjection} summarizes our feature injection strategies.
The first column lists the program properties we are going to inject,
including machine instructions and loading the address of a symbol or data.
The second and third columns 
list the modification primitives needed to inject features 
that can be extracted by NDISASM and radare2.
A cell with ``NA'' means that the binary analysis tool cannot extract the program property.
We discuss the non-NA cells in more details:

\begin{itemize}
\item Instructions extracted by NDISASM: The modification primitive \code{InsertNonCodeBytes(I)} 
creates a new non-loadable section in the binary to store the bytes representing new instructions. 
As NDISASM disassembles all bytes in the target binary, 
\code{InsertNonCodeBytes(I)} ensures that the features are injected and the functionality is unchanged.

\item Machine instructions extracted by radare2: The modification primitive \code{InsertFunction(I)} creates
a new function in which we store the inserted instructions. 
To ensure that radare2 disassembles the inserted code,
\code{InsertFunction(I)} creates a new code section to store the inserted instructions,
appends a return instruction at the end,
and create a new function symbol to point to the inserted instructions.

\item Loading symbol \code{S}: The modification primitive \code{InsertSymbol(S)} 
inserts the symbol \code{S} into the target binary
and returns the address pointing to the symbol. 
It is important to properly fill in all fields of the symbol in the symbol table, 
including symbol type, symbol visibility, and symbol section index.
Binary analysis tools may ignore incomplete symbols, causing the injection to fail. 
We then use \code{InsertFunction(I)} to insert code that loads the address of the new symbol.

\item Loading data \code{D}: The modification primitive \code{InsertData(D)} 
inserts the specified data into the target binary.
We typically need to create a new data section to hold the injected data.
Then, we use \code{InsertFunction(I)} to insert code that loads the data.

\item Calling function \code{F}: The modification primitive \code{InsertCall(F)} 
inserts the specified function $F$ into the target binary, 
where $F$ can be a function from an external library.
In such case, we also need to add  information for dynamic linking into the target binary,
including a dynamic function symbol, a relocation entry, and a procedural linkage stub (PLT)
for performing the external call.
Then, we use \code{InsertFunction(I)} to insert code that calls $F$.
\end{itemize}

\subsubsection{Feature removal strategies}

\begin{table}[t!]
\small
    \centering
    \caption[Summary of feature removal strategies]
    {\small \textbf{Summary of feature removal strategies}.
    The first column lists the program properties to remove or replace.
    The second and third columns list the binary modification primitives for feature removal.}
    \begin{tabular}{ p{2.5cm} | c | c } %p{100mm} |}
    
    \cellcolor{gray!20}\textbf{Program Property}            &   \cellcolor{gray!20}\textbf{NDISASM}     &    \cellcolor{gray!20}\textbf{radare2} \\
    \hline
    Instructions \code{I} from debug sections  &  \code{Overwrite(I)}   &     NA \\
    \hline
    Instructions \code{I} from code sections  &  \multicolumn{2}{c}{\code{Swap(I)} or \code{InsertNop(I)} } \\
    \hline
    Addressing loading of symbol \code{S} &  NA    &  \code{SplitAddrLoad(S)} \\
    \hline
    Addressing loading of data \code{D}    &  NA    &  \code{SplitAddrLoad(D)} \\
    \hline
    Function call to function symbol \code{S} &  NA  & \code{ConvToIndCall(S)} \\
    
    \end{tabular}
    \label{table:featremoval}
\end{table}

Table~\ref{table:featremoval} summarizes our feature removal strategies.
The first column lists the program properties we are going to remove or replace.
The second and third columns list the binary modification primitives needed 
for removing a feature group:

\begin{itemize}
\item Instructions \code{I} from debug-time sections:
The modification primitive \code{Overwrite(I)} overwrites 
the target instruction bytes to other bytes.
This strategy does not change the program's functionality as debug-time sections are not used at link-time or run-time.

\item Instructions \code{I} from code sections: 
We design two strategies for this feature group. 
The modification primitive \code{Swap(I)} 
checks the operand dependencies and reorders the instructions if there is no dependency.
The modification primitive \code{InsertNop(I)} inserts a
nop instruction between the original instructions. 
Note that to insert a nop instruction, 
we may need to relocate the original instructions to a different location to create extra space for the nop.
Therefore, we prefer \code{Swap} over \code{InsertNop} if possible.

\item Addressing loading of \code{S}: 
The modification primitive \code{SplitAddrLoad(S)} splits the address loading instruction into two instructions
so that radare2 will not recognize the address loading.
We use the following two instructions: 
loading the address minus one into the target register and 
incrementing the target register.

\item Function call to \code{S}:
The modification primitive \code{ConvToIndCall(S)} converts a function call to \code{S} to 
an indirect (pointer-based) function call, so that radare2 will not recognize the call target.
\code{ConvToIndCall(S)} uses \code{SplitAddrLoad(S)} to load the function call target
and then generates an indirect call. 
Note that we need to save and restore the register used for performing the indirect call
if it is live at this point in the code.
\end{itemize}

\subsubsection{Deciding which strategy to apply}

We have several criteria to determine which strategy to use for a modified feature.
Based on the sign of $\delta_i$, we decide whether we need to inject (see Table~\ref{table:featinjection})
or remove (see Table~\ref{table:featremoval}) features.
Based on the address where the feature was extracted,
we determine from which section the feature is extracted, 
including debug-time sections, code sections, or data sections.

For features extracted from code sections,
we determine whether the feature describes 
a function call, loading a symbol, or loading data. 
If none of the three cases applies, 
the feature describes just instructions, 
and no other program property needs to be modified.

\subsubsection{Generating diverse adversarial binaries}

Table~\ref{table:featinjection} and Table~\ref{table:featremoval} show 
one set of feasible modification strategies to inject and remove features,
out of a large space for binary modification.
Other modification strategies can be designed to achieve the same goals
of feature injection and removal.
We use two examples to show other possibilities of binary modification.
Attackers can add more modification strategies to
add diversity to the attack.

\textbf{Use randomization:} Several of our modification strategies can incorporate randomization
to generate diverse adversarial binaries. 
\code{Overwrite(I)} overwrites 
the target instruction bytes to other bytes. Here, we can randomly generate the overwritten bytes.
Similarly, \code{SplitAddrLoad(S)} can split the address 
loading instruction into two instructions with randomization: 
loading the address a randomly generate integer into the target register and 
incrementing the target register with generated integer.

\textbf{Generate semantically equivalent instructions:} 
The natural way to remove instruction features is to replace existing instructions
with semantically equivalent instructions.
Superoptimizer fits our goal here \cite{Massalin1987SLS},
which takes machine instructions as input, and outputs machine instructions
that compute the same functionality as the input.
It is expensive to perform superoptimization in a general case.
However, as we typically need to replace only short instruction sequences,
the search space would be relatively small.
Therefore, superoptimization is a promising method for generating semantically equivalent instructions.

\section{Evaluations}
\label{sec:evasion:eval}

We evaluate several aspects of our attacks:
(1) whether we can effectively perform untargeted attacks to evade authorship identification,
(2) whether we can effectively perform targeted attacks to impersonate someone else,
(3) how effective are different feature vector modification strategies,
including random modification, the CW attack, and the PGD attack,
(4) which features are modified in our attacks and which binary modification strategies are commonly used,
(5) whether our post-processing steps are effective for reducing the number of modified features,
and (6) why some of our attacks failed.
Our evaluations show that  

\begin{itemize}
\item Our untargeted attacks are effective. We achieved 96\% success rate in our experiments,
showing that we can effectively suppress authorship signals.
\item The success rate of our targeted attacks are 46\% on average, showing
that it is significantly more difficult to impersonate someone else.
\item The key to effective feature modification 
is utilizing feature correlation information. 
With feature correlation information, even random modification can achieve
decent success rate for untargeted attacks. 
Not surprisingly, adversarial learning based strategies such as the CW attack
perform significantly better than random modification on untargeted attacks,
and are necessary for targeted attacks.
\item The top modified features describe function calls.
This indicates that authorship identification classifiers heavily rely on function calls
to identify authors. Therefore, inserting function calls that are associated with other authors
is an effectively way to cause misprediction.
\item Without our post-processing, 
there are 80 features to modify on average.
With our post-processing,
there are only 10 features to modify on average.
Therefore, our post-processing procedure can significantly reduce the 
number of changed features for launching a successful attack.
\item For failed untargeted attacks, the lack of strategies for
modifying CFG features and decompiled source features is the reason for failure.
For failed targeted attacks,
about a third of the cases are caused by lack of modification strategies for 
CFG and decompiled source features;
the other two thirds of the cases failed because the targeted CW attack cannot generate
a feature vector that both corresponds to a real binary and causes the required misprediction.
\end{itemize}

\subsection{Evaluation Methodology}

We evaluated our techniques by attacking classifiers 
trained with the techniques presented by Caliskan-Islam et al. \cite{caliskan2018coding}
(described in Section~\ref{sec:example:background}).
Our experiments consist of the following steps:

\begin{enumerate}
\item Randomly sample $K$ authors from the Google Code Jam data set 
of around 1000 authors
used by Caliskan-Islam et al. \cite{caliskan2018coding}.
This data set consists of the source code of single-author programs,
each with an author label.
\item Compile all the programs written by the sampled authors 
with GCC 5.4.0 and -O0 optimization.
Each author had an average of 8 binary programs.  
\item Split the binaries into a training set and a testing set, 
with a size ratio of about 7:1.
\item Train a random forest classifier with the training set.
\item Perform our attack on each binary in the testing set for 
which the target classifier makes the correct prediction. 
For each test binary, we perform one untargeted attack, and $K-1$ targeted attacks.
The targeted attacks attempt to cause misprediction for each of the incorrect authors.
\end{enumerate}

We varied $K$ from 5 to 100 to investigate 
how the number of training authors impact the effectiveness of our attacks.
For each value of $K$, we repeated the experiments five times and report the averaged results.
We used Scikit-learn \cite{scikit-learn} for training random forest classifiers,
Tensorflow \cite{TensorFlow} for training substitute classifiers,
and Dyninst \cite{DyninstAPI} for implementing our binary modification strategies.

We implemented the CW attack, the PGD attack and two versions of
random modification for feature vector modification.
The two random strategies are
\code{rand}, which randomly modifies features without considering the feature correlation,
and \code{rand-cor}, which uses the feature correlation information derived in Section~\ref{sec:framework:corr}.
The CW attack and the PGD attack 
are extended to use the feature correlation information.
We used the $L_0$ version of the CW attack 
because it is designed to minimize the number of modified features.
In all our experiments, 
we keep the features unmodified,
for which we do not have corresponding binary modification strategies. 
We use \code{rand} as the baseline,
\code{rand-cor} to investigate the importance of feature correlation,
and use the CW attack and the PGD attack to investigate 
how much gain the adversarial learning based strategies can provide.

In our experiments, the CW attack on average modifies 10 features 
and generates fewer than 200 feature vectors before finding a desired feature vector.
For fair comparison, 
\code{rand} and \code{rand-cor} randomly choose 2 to 20 features to modify
and are allowed to repeat 5000 times to find a desired feature vector.

We use success rate to measure the effectiveness of our attacks, defined as
\begin{equation}
\frac{\#\ of\ successful\ attacks}{\#\ of\ total\ attacks}
\end{equation}

An attack is successful if 
the binary generated by our attack 
caused the target classifier to make an incorrect prediction.
For untargeted attacks,
incorrect prediction means any of the incorrect authors.
For targeted attacks,
incorrect prediction means the specific targeted author.
We measure the combined success rate of 
all the implemented feature vector modification strategies
and also their individual success rates. 

\subsection{Evaluation Results}

\begin{table*}
   \small
   \centering
   \caption[Evaluation results of our attacks]{\small \textbf{Evaluation results.} The second and the third columns
   list the accuracy of the target classifiers and the substitute classifiers.
   The fourth and the fifth columns show the combined success rate for untargeted and targeted attacks.}
   \begin{tabular}{ r | r r | r  |  r }
   \cellcolor{gray!20}& \cellcolor{gray!20} & \cellcolor{gray!20}&\cellcolor{gray!20} &\cellcolor{gray!20} \\
   \multirow{-2}{*}{\cellcolor{gray!20}$K$} & 
   \multirow{-2}{2.5cm}{\raggedright \cellcolor{gray!20}Target classifier accuracy} &
   \multirow{-2}{2.7cm}{\raggedright \cellcolor{gray!20}Substitute classifier accuracy} & 
   \multirow{-2}{2.5cm}{\raggedright \cellcolor{gray!20}Untargeted attack success rate} & 
   \multirow{-2}{2.2cm}{\raggedright \cellcolor{gray!20}Targeted attack success rate} \\
   
      \hline
5   &  100\%  & 100\%  &  88\% & 88\% \\
15   & 100\%  & 80\%  &  93\% & 51\% \\
30   &  89\%  & 73\%  &  98\%  & 47\% \\
50   &  86\%  & 69\%  &  100\% &  31\% \\
100   &  82\%  & 68\%  &  100\% &  14\% \\
   \hline
Average   &     91\%  & 78\%  &  96\% &  46\% \\

  \end{tabular}
   \label{table:results}
\end{table*}

The first question to answer in our evaluation is how effective is our attack.
The results are shown in Table~\ref{table:results}.
In this table, the second and the third columns
are the accuracy of the target classifiers and the substitute classifiers.
The fourth and the fifth columns list the 
the combined success rate of untargeted and targeted attacks.
Our untargeted attack has a \textit{96\% success rate on average,
showing that we can effectively suppress authorship signal}.
However, our targeted attacks did not enjoy the same success as the untargeted ones.
Our targeted attack has a 46\% success rate on average,
showing that it is significantly more difficult to impersonate a specific programmer's style.

Table~\ref{table:results} also shows 
how the number of training authors $K$ impacts the effectiveness of our attacks.
For untargeted attacks,
our success rate increases as $K$ increases.
Untargeted attacks only need to cause misprediction
against any of the $K-1$ incorrect authors. 
The larger the $K$, the more incorrect authors our attacks can work with,
and the higher the success rate.
For targeted attacks,
our success rate decreases as $K$ increases.
Targeted attacks must cause misprediction against a specific target author. 
The larger the $K$, 
the more non-target authors our target attack must avoid,
and the more difficult the targeted attack. 

The accuracy gap between the target classifier
and the substitute classifier does not obviously impact the success rate of our attack.
As shown in Table~\ref{table:results},
The accuracy gap ranges from 0\% to 20\%.
The success rates of both untargeted and targeted attacks 
do not exhibit an obvious correlation with the accuracy gap.

We then investigate the effectiveness of each feature vector modification strategies.
Figure~\ref{fig:untargetd} shows the results for untargeted attacks.
The baseline \code{rand} did not perform well 
because many generated feature vectors violates the feature correlation
and thus do not correspond to real binaries.
\code{rand-cor} performed much better than \code{rand},
showing that it is crucial to capture feature correlation.
The CW and the PGD attacks performed the best, 
showing that adversarial learning based strategies are effective 
for exploring the large feature space to find required feature vectors.

\begin{figure}
   \centering
   \setlength\figureheight{5cm} 
   \setlength\figurewidth{6cm}

   \begin{tikzpicture}

\begin{axis}[%
width=\figurewidth,
height=\figureheight,
scale only axis,
xmin=0, xmax=100,
xlabel={The number of training authors},
ymin=0, ymax=100,
ylabel={Success rate \%},
axis lines*=left,
legend pos=south east,
legend style={draw=black,fill=white,legend cell align=left, font=\small}]
\addplot [
color=red,
mark=otimes*,
mark options={solid},
mark size=1.5pt,
dashed
]
table{
5 82
15 86
30 96
50 100
100 99
};
\addlegendentry{CW};
\addplot [
color=blue,
mark=triangle*,
mark options={solid},
mark size=1.5pt,
dashed
]
table{
5 88
15 86
30 98
50 98
100 97
};
\addlegendentry{PGD};
\addplot [
color=black,
mark=square*,
mark options={solid},
mark size=1.5pt,
dashed
]
table{
5 71
15 93
30 61
50 98
100 94
};
\addlegendentry{\code{rand-cor}};

\addplot [
color=black,
mark=*,
mark options={solid},
mark size=1.5pt,
dashed
]
table{
5 0
15 7
30 22
50 50
100 71
};
\addlegendentry{\code{rand}};

\end{axis}
\end{tikzpicture}%
 
   \caption{\small \textbf{Comparison of feature vector modification strategies for untargeted attacks} The x-axis is the number of training authors. The y-axis is the success rate.}
   \label{fig:untargetd}
\end{figure}

Figure~\ref{fig:targetd} shows the results for targeted attacks.
Randomization based strategies have little success in targeted attacks,
showing that adversarial learning based techniques are necessary for targeted attacks. 
The PGD attack is designed for only untargeted attack, so we did not include it in this experiment.

\begin{figure}
   \centering
   \setlength\figureheight{5cm} 
   \setlength\figurewidth{6cm}

   \begin{tikzpicture}

\begin{axis}[%
width=\figurewidth,
height=\figureheight,
scale only axis,
xmin=0, xmax=100,
xlabel={The number of training authors},
ymin=0, ymax=100,
ylabel={Success rate \%},
axis lines*=left,
legend pos=north east,
legend style={draw=black,fill=white,legend cell align=left, font=\small}]
\addplot [
color=red,
mark=otimes*,
mark options={solid},
mark size=1.5pt,
dashed
]
table{
5 88
15 51
30 47
50 31
100 14
};
\addlegendentry{CW};
\addplot [
color=black,
mark=square*,
mark options={solid},
mark size=1.5pt,
dashed
]
table{
5 49
15 13
30 7
50 2
100 1
};
\addlegendentry{\code{rand-cor}};

\addplot [
color=black,
mark=*,
mark options={solid},
mark size=1.5pt,
dashed
]
table{
5 0
15 0
30 0
50 0
100 0
};
\addlegendentry{\code{rand}};

\end{axis}
\end{tikzpicture}%
 
   \caption{\small \textbf{Comparison of feature vector modification strategies for targeted attacks} The x-axis is the number of training authors. The y-axis is the success rate.}
   \label{fig:targetd}
\end{figure}

We then investigate what are the commonly used binary modification strategies and 
what are the commonly modified features.
In Table~\ref{table:mod}, we list the number of times that a modification primitive is used
in our untargeted attacks for $K=30$.
The most frequently used primitive is \code{InsertCall},
indicating that the target classifiers 
heavily rely on function call features to identify authors.
So, inserting function calls that are associated with other authors
is an effectively way to cause misprediction.
\code{SplitAddrLoad} ranks second,
showing that the target classifiers also rely on features that describe the loading of a symbol to identify authors.
\code{InsertFunction} ranks third,
showing that inserting instructions that are typically seen in programs written by other authors
is also effective for causing misprediction.
\code{Swap} and \code{InsertNop} serve the purpose of removing instruction features.
These two primitives have an effectiveness similar to \code{InsertFunction},
indicating that removing distinct instruction sequences associated 
with an author is effective for causing misprediction.
Other strategies including editing debug sections, inserting data, and inserting symbols,
all play important roles in our attacks.

\begin{table}
   \centering
   \caption[Binary modification primitive used]
   {\small \textbf{Number of times that a binary modification primitive is used in untargeted attacks}.
   The numbers are from the attacks for 30 training authors ($K=30$).
   The rows are sorted in a decreasing order.}
   \begin{tabular}{l | r}
   Modification primitive & Times used \\
   \hline
   \code{InsertCall}  & 586 \\
   \code{SplitAddrLoad} & 292 \\
   \code{InsertFunction} & 196 \\
   \code{Swap \& InsertNop} & 193 \\
   \code{ConvToIndCall} & 115 \\
   \code{InsertNonCodeByte} & 102 \\
   \code{Overwrite} & 85 \\
   \code{InsertData} & 68 \\
   \code{InsertSymbol} & 45 \\
  \end{tabular}
   \label{table:mod}
\end{table}

Next, we investigate how many features we need to change to cause misprediction.
In Table~\ref{table:featchange}, 
the second column shows the number of changed features generated by the untargeted $L_0$ CW attack,
and the third column shows the number of changed features after our post-processing step.
Before post-processing, 
there are 80 features to modify on average.
After the post-processing,
there are only 10 features to modify on average.
Our results show that our post-processing procedure can significantly reduce the 
number of changed features for performing a successful attack.

\begin{table}
   \centering
   \caption[The number of feature changed by our untargeted attacks]
   {\small \textbf{The number of feature changed by our untargeted attacks.} The second column
   lists the average number of features changed by the $L_0$ CW attack. The third column shows the average number
   of features changed after our post-processing.}   
   \begin{tabular}{r | r r}
    $K$       &  $L_0$ CW attack  &  Our post-processing \\ 
   \hline
5   &  57  &  9 \\
15   &  107   &  11 \\
30   &  87   &  11 \\
50   &  59   &  9 \\
100   &  92   &  11  \\
   \hline
Average    &  80   &  10 \\
   \end{tabular}
   \label{table:featchange}
\end{table}

\subsection{Analysis of Failed Attacks}
\label{sec:eval:analysis}
Our attack contains two key steps:
feature vector modification to generate a vector that
both corresponds to a real binary and causes the required misprediction,
and input binary modification to generate a new binary that matches the adversarial feature vector.
A failure in either of the two steps would lead to a failed attack.
Feature vector modification fails
when it cannot find such an adversarial feature vector that 
corresponds to a real binary and 
causes the required misprediction.
Input binary modification fails when
it does not generate a new binary that causes the required misprediction.
We found that
feature vector modification accounts for all the failed attacks.

We break down the reasons of why our feature vector modification
step would fail to generate an adversarial feature vector.
Recall that our feature vector modification includes randomization, the CW attack, and the PGD attack.
These modification strategies do not consider whether the generated vector would correspond to a real binary.
We adapted them in three ways to
generate vectors that correspond to a real binary.
First, 
as we implemented binary modification strategies for only instruction features,
the CFG features and decompiled source code features are not modified
during feature vector modification.
Second, as the value of an instruction feature represents the number of times that 
this feature appears in a binary,
the feature value is an integer.
However, the CW attack and the PGD attack do not guarantee to generate integer values.
So, we round the results of the CW attacks to the nearest integer values.
Third, we capture feature correlation and merge correlated features.
We can then divide failed feature vector modification into two categories:

\hangindent=1.5cm \textit{Lack of modification strategies for CFG and decompiled source features:}
It may not be sufficient
to modify only instruction features to evade authorship identification.  
Failed attacks in this category need 
binary modification strategies for CFG and decompiled source features.

\hangindent=1.5cm \textit{Insufficient handling of finding feature vectors corresponding to real binaries:}
Our techniques for generating feature vectors
that correspond to real binaries need further improvement. 
For example, we currently capture only linear correlation between features.

We found that all the failed untargeted attacks were due to 
not being able to modify CFG or decompiled source features.
For failed targeted attacks,
not being able to modify CFG or decompiled source features
explained about 34\% of the failed cases;
not being able to find a feature vector that corresponds to a real binary
explained the other failed cases.
Our analysis shows that to improve untargeted attacks, we
need to continue to design new modification strategies for CFG and decompiled source features.
To improve targeted attacks, we also need to 
improve the targeted CW attack to find feature vectors that correspond to real binaries.

While our binary modification strategies were able to match all modified features,
We found that they sometimes caused side effects
and changed features that should not have been changed.
Fortunately, such side effects did not impact the prediction results.
The number of unintended changes ranged from 0 to 20. 
Most of the unintended changes were made to NDISASM instruction features.
This is because 
our feature injection strategies
often insert new code and data sections,
which in turn requires changes to the program header of the binary.
As NIDSASM disassembles all the bytes in the binary, 
the changes in the program header would cause unintended changes to NDISASM instruction features.
It is not surprising that such unintended changes did not impact the prediction results
as the program header is unlikely to carry authorship signals.

In summary, the our evaluations show that our attack framework is effective for untargeted attacks
and we can practically suppress authorship signals.
Performing targeted attacks is significantly more difficult
than untargeted attacks.
Our results also reveal weaknesses in current authorship identification techniques.
Many features used in current authorship identification techniques 
are based on program properties that are easy to fabricate, 
such as function calls and symbols.
We have shown that we can automatically modify these features, 
making such classifiers vulnerable to test time attacks.

\section{Discussion and Related Work}
\label{sec:discussion}

To the best of our knowledge, we are the first project
to perform binary code authorship evasion.
In this section,
we place our work in a broader context
and discuss several related research areas.

\textbf{Stealthy binary rewriting:} 
Stealthiness is an important goal of out attack.
We observed that the generated binaries show distinct characteristics 
such as the presence of additional code sections and 
springboard jump instructions
from original code sections to newly added code sections.
These distinct characteristics are introduced by Dyninst
and every binary modified by Dyninst exhibits such characteristics.
Since Dyninst is also widely used in many benign applications,
such as binary hardening techniques 
\cite{vanderVeen2016ATC,vanderVeen2015PCC,pawlowski2017marx}.
The presence of Dyninst footprints does not necessarily indicate
the presence of tamperers. 

We are aware of other binary rewriting techniques, 
such as reassembly \cite{Wang2015RD, wang2017ramblr}.
Reassembly disassembles the binary and creates artificial symbols for data and code references.
Binary rewriting is performed by first modifying the assembly code
and then re-assembling the code.
Reassembly has the advantage that 
code can be injected or removed in place,
thus providing better stealthiness.

We chose to use Dyninst for binary rewriting as it is a mature
and widely used tool. We leave the exploration of using reassembly 
for binary rewriting as future work.

\textbf{Adversarial learning on malware detection:}
While our work is not directly targeted to malware detection,
we believe our techniques can contribute to this field in two ways.

First, a common threat model of adversarial learning on malware detection
assumes that attackers can only inject features and cannot remove features.
Monotonic classification \cite{Incer2018ARM} ensures 
that an adversary will not be able to evade the classifier by adding more features.
Our results show that we can effectively remove features,
challenging the validity of their threat model.

Second, existing techniques for adversarial learning on malware detection 
have focused on generating adversarial feature vectors to cause misprediction,
but have not focused on generating new binaries that match their feature vectors.
We show that it is possible to perform end-to-end attack by
generating new binaries.

\textbf{Evading source code authorship:} 
Simko et al. \cite{simko2018recognizing} performed a study
of evading source code single-author identification.
28 programmers participated in their study, 
including undergrad students, former or current software developers.
Each programmer
was given code from author X and Y  
and then was asked to modify source code written by X 
to look like code written by Y.
Essentially, the study performed manual attacks.

Simko et al. found that effective source code changes
are mostly local, involving a few lines of code
For example, it is effective to copy entire lines of code written by Y into code written by X,
modify variables names, add or remove macors, and change newlines and spaces between operators.
The modification strategies presented in this study 
are unlikely to achieve equal success for evading 
binary code authorship identification,
as many of the modifications are irrelevant at the binary code level, 
such as typographic changes, variables renaming, and modifying macros.

\section{Conclusion}
We have presented our attack framework for performing authorship evasion.
Our attack framework includes components for analyzing feature correlation,
generating feature vectors to cause misprediction,
and binary modification strategies to match the generated feature vectors.
Our evaluations have shown that 
our attack framework is effective for untargeted attacks,
which is to cause misprediction to any of the incorrect authors.
Targeted attacks are significantly more difficult to achieve,
which is to cause misprediction to a specific one among the incorrect authors.

Our attack experiences show that it is not secure to rely on features derived from
program properties that are easy to modify, 
such as function calls, symbols, data, and instructions.
Authorship identification techniques must consider the trustworthiness of the features.

%\section{Acknowledgments}
\setstretch{1}
\small
\bibliographystyle{abbrv}
\bibliography{adversarial}

\end{document}